\documentclass[useAMS,usenatbib,usegraphicx]{mn2e}
\def\X{{\mathrm{x}}}
\def\Y{{\mathrm{y}}}

\def\n{{\rm n}}

\def\p{{\rm p}}

\def\tr{\tilde{r}}

\newcommand{\be}{\begin{equation}}
\newcommand{\ee}{\end{equation}}
\newcommand{\beq}{\begin{eqnarray}}
\newcommand{\eeq}{\end{eqnarray}}
\newcommand{\bear}{\begin{eqnarray}}
\newcommand{\eear}{\end{eqnarray}}
\newcommand{\ba}{\begin{array}}
\newcommand{\ea}{\end{array}}

\title[Modelling pulsar glitches]{Modelling pulsar glitches with realistic pinning forces: a hydrodynamical approach}

\author[B.Haskell et al.] {B.~Haskell$^{1,2}$, P.~M.~Pizzochero$^{3,4}$, T.~Sidery$^{5,6}$\\$^1$School of Mathematics, University of Southampton, Southampton SO17 1BJ, UK\\$^2$Astronomical Institute ``Anton Pannekoek'', University of Amsterdam, Science Park 904, 1098 XH Amsterdam, Netherlands\\$^3$Dipartimento di Fisica, Universit\`a degli Studi di Milano,Via Celoria 16, 20133 Milano, Italy\\$^4$Istituto Nazionale di Fisica Nucleare, Sezione di Milano, Via Celoria 16, 20133 Milano, Italy\\$^5$FENS, Sabanci University, Orhanli, 34956 Istanbul, Turkey\\$^6$School of Physics and Astronomy, University of Birmingham, Edgbaston, Birmingham, B15 2TT, UK}
\begin{document}
\maketitle
\begin{abstract}
Although pulsars are one of the most stable clocks in the universe, many of them are observed to 'glitch', i.e. to suddenly increase their spin frequency $\nu$ with fractional increases that range from  $\frac{\Delta\nu}{\nu}\approx 10^{-11}$ to $10^{-5}$. In this paper we focus on the 'giant' glitches, i.e. glitches with fractional increases in the spin rate of the order of $\frac{\Delta\nu}{\nu}\approx 10^{-6}$, that are observed in a sub class of pulsars including the Vela. We show that giant glitches can be modelled with a two-fluid hydrodynamical approach. The model is based on the formalism for superfluid neutron stars of Andersson and Comer (2006) and on the realistic pinning forces of Grill and Pizzochero (2011). We show that all stages of Vela glitches, from the rise to the post-glitch relaxation, can be reproduced with a set of physically reasonable parameters and that the sizes and waiting times between giant glitches in other pulsars are also consistent with our model. 
\end{abstract}

\section{Introduction}

The timing of radio pulsars provide us with one of the most stable clocks in the universe. Many pulsars exhibit, however, sudden increases in the spin rate, known as 'glitches'. To date there have been glitches reported in more than 100 pulsars\footnote{http://www.jb.man.ac.uk/pulsar/glitches/gTable.html} \citep{Espinoza}, with fractional jumps in the spin rate of $\frac{\Delta\nu}{\nu}\approx 10^{-11}$ to $10^{-5}$.

There is still no clear consensus on the origin of these phenomena, but it is thought that glitches (at least the giant glitches that are observed in somewhat older, colder pulsars such as the Vela) are due to angular momentum being stored in a superfluid component of the star that is temporarily decoupled from the charged component to which the electromagnetic emission is anchored. When the two components recouple there is a sudden transfer of angular momentum to the crust, which gives rise to the observed spinup \citep{Anderson}.
A superfluid rotates by forming a quantised array of vortices, the distribution of which determines the rotational profile of the star. The main idea at the base of most glitch models is that vortices can {\it{pin}} to the crustal lattice \citep{Anderson,Alpar77,Pines80,Alp81,Anderson82}, allowing for a lag to build up between the superfluid and the charged component. 

Following the seminal work of \citet{BPP69} several models have been developed to explain the dynamical evolution of the pinned superfluid coupled to the crust, mainly to with the intent of explaining the observed timescales of days to months in the post-glitch recovery of the Vela pulsar.
Essentially one can distinguish two classes of models, those that assume that the relaxation is due to the weak coupling between the superfluid and the crust due to the interaction between free vortices and the coulomb lattice of nuclei \citep{Jones2, JonesGL, JonesGL2} and those which rely on the creep model based on the seminal work of  \citet{Alpar1} and later adaptations \citep{ALS,LEB}. In this case the assumption is that as a lag develops between the crust and the superfluid, vortices {\it creep} through the crust at a rate that is highly temperature dependent, gradually transferring angular momentum and leading to a steady state spindown. However, when the lag approaches a critical value, it is possible for an instability to trigger a catastrophic unpinning event associated with a sudden transfer of angular momentum to the crust, after which vortices repin and a new equilibrium is reached.
These models work remarkably well in explaining the post-glitch relaxation, but struggle to explain the exact nature of the perturbation which triggers it. Such an event may be triggered by large temperature perturbations \cite{LE96}, caused for example by starquakes \cite{BP71,Cheng92}, the interactions of the proton vortices and the crustal magnetic field \citep{SC99}  or, in the presence of strong crustal pinning, a two-stream instability in the superfluid flow may develop, leading to a sudden transfer of angular momentum \citep{kglitch} 
Recently, however,  \citet{NewM} have shown that glitches can be triggered by unpinning avalanches of the vortex array, leading to very promising results for the overall glitch size and waiting time distributions (see also \citet{Melatos1, Melatos2}).

An entirely different mechanism was proposed by \citet{Rud91}, who suggested that  vortices pinned to the crust stress it to the point of fracture and then move outwards with the matter they are pinned to. Several other ``starquake'' models have been proposed \citep{EL2000,FLE2000} but they are all based on the main assumption that a starquake triggers the outward motion of vortices pinned to the matter and the local rearrangement of the crust then causes an increase of the angle between the rotation axis and the magnetic axis, leading to an increased electromagnetic braking torque on the star. Note that starquakes had been invoked as a possible explanation for pulsar glitches very soon after the first observations, but had been ruled out as they could not account for the giant glitches of the Vela \citep{BP71,Rud75}.  This mechanism may, however, play a role in other systems and in fact there is some evidence that smaller glitches in younger and hotter pulsars such as the Crab may be linked to starquakes \citep{Mid}. Furthermore it has been suggested that the glitch may not be exclusively linked to vortex dynamics in the crust, but that the interaction between the vortices and the superconducting flux tubes in the core may contribute to the event \citep{Rud98,trevnew}.

One of the main difficulties in constructing a glitch model is, however, the relative lack of realistic calculations for the pinning forces. Several calculations exist of the pinning {\it energy} and of the force per pinning site \cite{Alpar77, Alpar1,EB88,LE91, PV,PBJ98, DP03, DP04, DP06,Avo} but such calculations of the pinning force neglect the finite length of the vortices and rely on simplified geometries. Recently, however, \citet{Linkn} has analysed the motion of a vortex in a random potential and, even more crucially, \citet{pinning1} have obtained realistic estimates of the pinning force {\it per unit length} of the vortex, the quantity that can be directly compared to the Magnus force which tends to push the vortices out and force the superfluid into corotation with the crust.

\citet{Pierre1} has recently shown that a simple analytic model, based on the catastrophic unpinning paradigm and the realistic pinning forces calculated by \citet{pinning1} (see also \citet{pinning1phd}), can describe glitches in the 'Vela' like pulsars, that are typically older, exhibit larger glitches ($\frac{\Delta\nu}{\nu}\approx 10^{-7}$ to $10^{-6}$) and always show a decrease in the frequency derivative after the glitch \citep{Espinoza}. Essentially the assumption is that as the star spins down the vorticity is accumulated in the strong pinning region and then released once the resulting Magnus force exceeds the maximum of the pinning force.

In order to make quantitative predictions for the observed jump in frequency and subsequent relaxation of the crust it is, however, also necessary to describe the evolution of the various fluid components that form the NS. In order to do this it is convenient to follow a hydrodynamical approach that does not deal with vortex motion directly, but rather follows the evolution of the separate fluid components. In particular we shall make use of the multifluid formalism developed by \citet{NilsandGreg} and recently applied, albeit in a much simpler formulation, to the study of pulsar glitches by \citet{glitch}. In the following we shall thus go beyond the analytic model of \citet{Pierre1} and show how the realistic pinning results of \citet{pinning1} can be incorporated into this formalism and used to accurately reproduce, for a reasonable range of parameters, the observed properties of pulsar glitches. 
Furthermore, as we will discuss in the following sections, we will assume that the relaxation is entirely due to the mutual friction between the superfluid component and the crust. In a sense we are thus taking the view of \citet{JonesGL} that the relaxation is due to the dynamics of vortices close to corotation with the superfluid, rather than to vortex 'creep'. Note, however, that the creep model has been shown to be consistent with the observed relaxation timescales of the Vela pulsar in the linear regime \citep{Alp89},  and we can thus qualitatively account for it by rescaling the mutual friction parameters, as will be discussed in the following. Finally let us also point out that an effect that is neglected in this preliminary model is that of Ekman pumping at the crust-core interface, which has been shown, together with shear viscosity, to be relevant in explaining the post-glitch relaxation times \citep{vE1} and should be included in future developments.

Let us stress that such detailed theoretical work is crucial at this time, as the new low frequency radio telescope LOFAR has just come online and begun observations of pulsars and fast radio transients \citep{LOFAR}. As development continues LOFAR, and in the future the Square Kilometer Array (SKA), are likely to not only provide a wealth of data on both known and new glitching pulsars, but also resolve (or at least tightly constrain) the glitch rise time, thus allowing us to test our theoretical understanding of the glitch mechanism \citep{SKA}. Furthermore a glitch may trigger a gravitational wave (GW) burst  which could be detectable by next generation GW observatories \citep{glitch,vE2}. By accurately determining the glitch time LOFAR and the SKA could be used to trigger a directed GW search. 

Finally the constructions of long-term hydrodynamical simulations of NS interior could also be extended to model not only large pulsar glitches, but more generally pulsar timing noise, in order to understand the role that the interior dynamics of the star plays compared to magnetospheric processes, which have been shown to be at work in some systems \citep{Lyne}. This is an issue that is of great importance for the current efforts to detects GWs with pulsar timing arrays \citep{Hobbs}.

\section{Two fluid equations of motion}

We model the pulsar as a two fluid system of superfluid neutrons and a charged component, which consists of the crust and the charged particles in the interior. As the neutrons are superfluid they will rotate by forming an array of quantised vortices, the interaction of which with the fluids will give rise to a weak coupling between the components, known as mutual friction.
To describe the equations of motion of the crust and of the superfluid we shall use the two-fluid formalism for neutron star cores developed by \citet{NilsandGreg}. We thus do not consider the equations of motion of the vortices, but rather model the macroscopic motion of two dynamical degrees of freedom, representing the superfluid neutrons and a neutral conglomerate of protons, electrons and all non-superfluid components that are strongly coupled to them. Assuming that the individual species are conserved, we have the
standard conservation laws
\be
\partial_t \rho_\X + \nabla_i (\rho_\X v_\X^i) = 0\label{continuo1}
\ee
where the constituent index x may be either p or n. The Euler equations take the form
\be
(\partial_t + v_\X^j \nabla_j ) (v^\X_i+\varepsilon_\X w^{\Y\X}_i) +\nabla_i (\tilde{\mu}_\X+\Phi)
+ \varepsilon_\X w^j_{\Y\X} \nabla_i v^\X_j= f^\X_i/\rho_\X
\label{Eulers}\ee
where $w_i^{\Y\X}=v_i^\Y-v_i^\X$ ($\Y\neq\X$), and $\tilde{\mu}_\X=\mu_\X/m_\X$ represents the
chemical potential (in the following we assume that $m_\p=m_\n$). Moreover, $\Phi$ represents the gravitational potential, and the parameter
$\varepsilon_\X$ encodes the  entrainment effect.
The force on the right-hand side
of (\ref{Eulers}) can be used to describe  other interactions, including dissipative terms \citep{NilsandGreg}. We will focus on the 
vortex-mediated mutual friction force. This means that we
consider a force of form \citep{trev}
\be
f^\X_i = \kappa n_v  \rho_\n \mathcal{B}' \epsilon_{ijk}\hat{\Omega}_\n^j w_{\X\Y}^k
+ \kappa n_v  \rho_\n  \mathcal{B} \epsilon_{ijk}\hat{\Omega}_\n^j \epsilon^{klm} \hat{\Omega}^\n_l w_m^{\X\Y}
\label{mf}\ee
where $\Omega^j$ is the angular frequency of the neutron fluid (a hat represents a unit vector), $\kappa=\frac{h}{2m_n}$ represents the quantum of circulation and $n_v$ is the vortex number per unit area. In particular the quantity $\kappa n_v$ is linked to the rotation rate of the neutrons and protons by the relation (where $\tilde{r}$ is the cylindrical radius):
\be
\kappa n_v=2\left[\Omega_\n+\varepsilon_\n(\Omega_\p-\Omega_\n)\right]+\tilde{r}\frac{\partial}{\partial\tilde{r}}\left[\Omega_\n+\varepsilon_\n(\Omega_\p-\Omega_\n)\right]\label{numero}
\ee

One can express the mutual friction force in terms of a dimensionless "drag" parameter $\mathcal R$
such that \citep{trev}
\be
\mathcal{B} = { \mathcal{R} \over 1 + \mathcal{R}^2} \ , \qquad \mbox{ and } \qquad
\mathcal{B}' = { \mathcal{R}^2 \over 1 + \mathcal{R}^2}
\ee
and is related to the usual drag parameter $\gamma$ by the relation
\be
\mathcal{R}=\frac{\gamma}{\kappa \rho_\n}
\ee
Note that for $\mathcal{R}\ll 1$ (which is relevant in our context) one has $\mathcal{B}\approx\mathcal{R}$ and $\mathcal{B}'\ll \mathcal{B}$.

In our model we shall consider the two components to be rotating around the same axis defined by $\hat{\Omega}_\p$. Furthermore the protons in the crust (to which the magnetic field is assumed to be anchored) are bound in ions which form a crystalline solid and are connected to the protons in the core on an Alfven crossing timescale, which for a neutron star core is of the order of a few seconds (or less if the core is superconducting) and thus much shorter than the dynamical timescales we are interested in (except possibly for the glitch rise time). Accounting for the elastic forces in the crust and Alfven waves in the core is clearly a prohibitive task, so as a first approximation we shall account for these effects by considering the proton conglomerate to be rigidly rotating.
We make no such assumptions for the neutrons which will, in fact, develop differential rotation if vortices migrate to different regions of the star, as can be seen from equation (\ref{numero}).

Following \citet{glitch} we can write the equations of motion for the angular frequency of the two components in the form:
\beq
\dot{\Omega}_\n(\tr)&=&\frac{Q(\tr)}{\rho_\n}\frac{1}{1-\varepsilon_\n-\varepsilon_\p}\label{eom1}\\
\dot{\Omega}_\p(\tr)&=&-\frac{Q(\tr)}{\rho_\p}\frac{1}{1-\varepsilon_\n-\varepsilon_\p}\label{eom2}-\frac{\mathcal{A}}{I_\p}\Omega_\p^3
\eeq
where we have defined
\be
Q(\tr)=\rho_\n\kappa n_v\mathcal{B}(\Omega_\p-\Omega_\n)
\ee
and
\be
\mathcal{A}=\frac{B^2\sin\theta^2 R^6}{6 c^3}
\ee
with $B$ the surface magnetic field strength, $\theta$ the inclination angle between the field and the rotation axis and $I_\p$ the moment of inertia of the proton fluid in the star. Note that the coupling between the components depends only on the dissipative mutual friction coefficient $\mathcal{B}$ and not on $\mathcal{B}'$.
If we now assume rigid rotation for the protons, i.e. a constant $\Omega_p$, we can can multiply equation (\ref{eom2}) by $\tr^2\rho_\n$ and integrate over the volume, thus recasting the system in the form
\beq
I_\p\dot{\Omega}_\p&=&-\frac{\mathcal{A}}\Omega_\p^3+\int \tilde{r}^2 Q(\tr) \frac{1}{1-\varepsilon_\n-\varepsilon_\p} dV\label{integrate1}\\
 \dot{\Omega}_\n(\tr)&=&\frac{Q(\tr)}{\rho_\n}\frac{1}{1-\varepsilon_\n-\varepsilon_\p}\label{integrate}
 \eeq

\section{The pinning force}
\label{pin}
Although vortex pinning in the crust has been suggested as the main mechanism behind pulsar glitches more than twenty years ago by \citet{Anderson}, until recently very little work has been devoted to obtaining realistic estimates of the pinning force. The main difficulty lies in the fact that, although there have been several estimates of the pinning energy and thus of the force per pinning site \citep{Alpar1}, it is in fact the pinning force {\it{per unit length}} acting on a vortex that is the relevant quantity to compare with the Magnus force if one is to understand when a vortex line can unpin \citep{Anderson82}.

In fact it has been argued by \citet{Jones1} that if one considers an infinitely long vortex line and averages over the various orientations with respect to the lattice, the pinning force will be negligible. Recent calculations by \citet{pinning1} have shown that for realistic configurations in which one considers a vortex to be rigid over lengthscales of 100-1000 Wigner-Seitz radii, the averaging process over mesoscopic lengthscales and different orientations of the lattice does indeed reduce the strength of the pinning force with respect to previous estimates based on the pinning force per pinning site. However, the values obtained in these calculations, that are in fact the first realistic estimate of the pinning force per unit length, are  still large enough to explain pulsar glitches \citep{Pierre1}. 

It is beyond the scope of the present work to make a comparison between  various superfluid gap models and equations of state (however see the discussion in \citet{Stefano}, who find that the main qualitative features of the model remain unchanged), as  at present we are mainly interested in understanding how to reproduce the main features of a glitch with a simplified large scale hydrodynamical neutron star model. Furthermore we shall use a Newtonian  model to describe the mutual friction dissipation between the two components, as at present the relativistic prescription for doing this is not fully developed. Given the inaccuracies inherent to a Newtonian calculation of a neutron star model any comparison between realistic equations of state would be at least dubious. Furthermore there are, as we shall see, huge uncertainties associated with the values of the mutual friction coupling parameters that would make such a calculation meaningless.

\begin{figure}
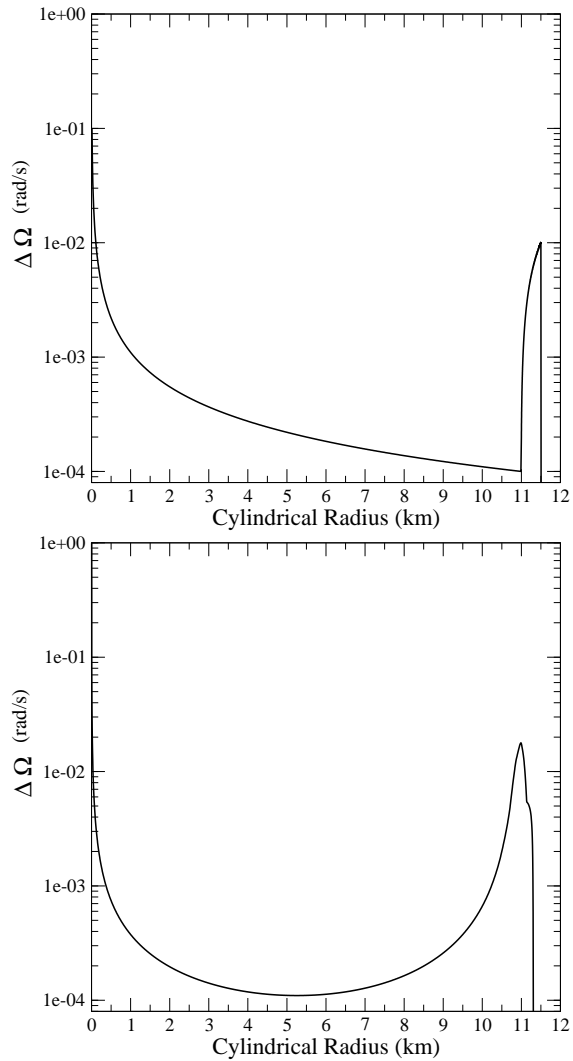

\includegraphics[height=7cm,clip]{unp2.eps}

\includegraphics[height=7cm,clip]{unp1.eps}
\caption{The top panel shows our approximation for the critical unpinning lag for a 12 Km, 1.4 M$_{\odot}$ star, modelled as an $n=1$ Newtonian polytrope. In the bottom panel we show for comparison the critical unpinning lag for a realistic model of a 12 km, 1.4 M$_{\odot}$ using the SLy equation of state \citep{SLy} as discussed in \citet{Stefano}.}\label{lag1}
\end{figure}

In this work we shall  focus on the case of continuous vortices  that thread the whole star, as in \citep{Pierre1}, and extract the main features of the calculation of \citet{Stefano} to construct a simplified model for the maximum lag that can be developed. 
In particular, as the maximum pinning force is obtained by balancing the Magnus force over a whole vortex the maximum lag between the two components ($\Delta\Omega$) will have cylindrical symmetry and be a function of the cylindrical radius $\tilde{r}=r\sin(\theta)$ only. Close to the rotation axis the radial behaviour is thus due to the Magnus force, as the vortices encounter a roughly similar number of pinning sites. The maximum lag thus falls off as $\approx 1/\tilde{r}$ until it reaches a minimum value of $\Delta\Omega\approx 10^{-4}$ rad/s at the base of the crust (note that in the analytic model of \citet{Stefano} the minimum is at somewhat higher densities, deeper in the outer core). The maximum lag then rises steeply in the equatorial region as the vortices are now completely contained in the pinning region and reaches a maximum value of $\Delta\Omega\approx 10^{-2}$ rad/s, in the inner crust. We model this behaviour as a linear rise of the lag between the minimum and maximum values. As we continue to move outward the lag then decreases linearly again as the vortices annihilate in the outer regions of the crust where the neutrons are no longer superfluid. A schematic representation of this is given in figure \ref{lag1}.
\begin{figure}
\includegraphics[height=7cm,clip]{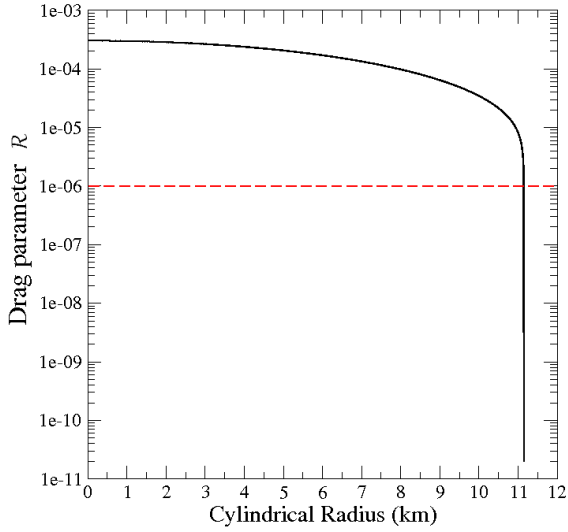}
\caption{The average drag parameter $\mathcal{R}$ as a function of cylindrical radius, obtained by averaging the core contribution $\mathcal{R}_c$ and crust contribution $\mathcal{R}_p$ over the length of a vortex. Note that we consider only the drag due to phonon excitation ($\mathcal{R}_p$) in the crust as we are assuming that Kelvin waves can only be excited once the critical unpinning lag has been exceeded. The horizontal line corresponds to $\mathcal{R}$ of the order of $10^{-6}$, which gives a coupling timescale of the order of 1 minute.}
\label{lag2} 
\end{figure} 

\section{Equation of state and drag parameters}

Let us discuss the fiducial NS model that we adopt for our calculations.
We consider a 1.4 $M_\odot$, with a radius $R=12$ km, and assume that the base of the pinning region is at $r=11.15$ km, at a density $\rho=9.6\times 10^{13}$, which is rougly consistent with the estimates of \citet{Stefano}. We take the background equation of state of our neutron star to be a $n=1$ polytrope, which allows us to make use of the explicit solution for the density profile
\be
\rho=\rho_c \frac{\sin\alpha}{\alpha}
\ee
with
\be
\alpha=\frac{\pi r}{R}\;\;\;\;\;\mbox{and}\;\;\;\;\;\rho_c=\frac{\pi M}{4 R^3}
\ee
This gives us a value $I_t=1.57\times 10^{45}$ for the total moment of inertia. However we need to consider the proton and neutron densities separately, so, following \citet{reis}, we shall consider a proton fraction $x_\p$ of the form:
\be
x_\p=\frac{\rho_\p}{\rho}=0.05\left(\frac{\rho}{10^{14}\mbox{g/cm$^3$}}\right)
\ee
where $\rho=\rho_\p+\rho_\n$ is the total density.

Finally in a fully consistent model one should be able to obtain the entrainment coefficients directly from the equation of state. There are however few realistic fully consistent equations of state for superfluid neutron star matter. Entrainment coefficients have, however, been obtained for the cores of stars containing hyperons \citep{Gus1,Gus2} and for the crust one can rely on calculations of the proton effective mass \citep{Chamel,Chamel2,Chamel3} which is related to our entrainment coefficients by the relation \citep{PCA01}:
\be
\varepsilon_\p=1-\frac{m^{*}_\p}{m_\p}
\ee
where $m_\p$ is the bare proton mass and $m^{*}_\p$ is the proton effective mass. Recent calculations \cite{Chamel} suggest that the proton effective mass is slightly lower than the bare mass in the core but can be larger in the crust. This means that the entrainment parameters will vanish close to the base of the crust \cite{Carter}. As this is the region we are mainly interested in for our evolutions and given the qualitative nature of this work, we shall take $\varepsilon_p=\varepsilon_n=0$.

The discussion of the mutual friction coefficients is more complex. The mechanisms that give rise to mutual friction are in fact different in different regions of the star. In the core we expect electron scattering of vortices to be at work, coupling neutrons and protons on a timescale of $\tau \approx 10$ rotational periods \citep{ALS, trev} (which for the Vela pulsar would give a timescale of slightly less than a second). Another possibility is that, if the protons are in a type II superconducting state, the interaction between fluxtubes and vortices will couple the two components on an even shorter timescale \citep{trevnew}, although if the flux tube tangle is sufficiently strong to "pin" the vortices this may lead to the opposite being true and the core being effectively decoupled for most of the evolution \citep{Rud98, Linkp}.  We shall discuss this possibility in more detail in the following.

In the crust, on the other hand, it has been shown by \citet{Jones3} (see also \citep{Jones2, Baym}) that coupling to sound waves in the lattice will be the main source of dissipation for low velocities of the vortices with respect to the lattice ($\le 10^2$ cm/s) while dissipation due to the excitation of Kelvin waves in the vortices will dominate for large relative velocities.

In table \ref{table1} we show the values of the drag parameters in the crust, for both these interactions, obtained from \citet{Jones2} by using recent values of the vortex-nucleus interaction \citep{DP06}. Unfortunately there is a large uncertainty associated with this calculation, as one must not only evaluate the energy dissipated by the interaction of the vortex with a single pinning site, but also sum coherently over the lengthscale associated with vortex rigidity (which \citet{pinning1} and \citet{pinning1phd} find to be of the order of $10^2-10^3$ Wigner-Seitz radii). The summation procedure leads to a poorly constrained reduction factor $\delta$, which \citep{Jones2} finds to be of the order $\delta\approx 10^{-4}$.
This is consistent with the results of \citet{pinning1}, who find that averaging over mesoscopic vortex length-scales and different crystal orientations leads to a reduction of up to two orders of magnitude in the pinning force per unit length, with respect from previous estimates that were obtained directly from the pinning force per pinning site. Given the scalings in the expressions for the drag parameters of \citet{Jones2}, this would lead to a reduction factor $\delta\approx 10^{-4}$ for the drag parameters. 
There is thus a large uncertainty associated with the values in table \ref{table1} and, furthermore, drag parameters $\mathcal{R}>>1$ are dubious as for such a strong interaction the perturbative treatment of \citet{Jones2} breaks down \citep{Baym}. This suggests that for our qualitative study we may take a constant drag parameter throughout the crust and study how the results depend on its variation.

\begin{table}
\caption{Drag parameters in the crustal regions defined by \citep{NegVaut}. In the last two lines we have applied a reduction factor $\delta=10^{-4}$, as described in the text. $\mathcal{R}_p$ indicates the drag parameter due to phonon excitations and $\mathcal{R}_k$ that due to Kelvon excitations. }
\begin{center}
\begin{tabular}{l l l l l l l l l l}
Zone&1& &2& &3& &4& &5\\
\hline
${\rho}/{10^{14}\mbox{g/cm$^3$}}$ &0.015& &0.096& &0.34& &0.78& &1.3\\
${\mathcal{R}_p}/{10^{-5}}$& 1.6& &0.7& &5.8& &0.025& &0.0\\
${\mathcal{R}_k}/{10^{2}}$&290& &5.8& &340& &0.085& &0.0\\
${\mathcal{B}_p}/{10^{-5}}$& 1.6& &0.7& &5.8& &0.025& &0.0\\
${\mathcal{B}_k}/{10^{-2}}$&0.003& &0.18& &0.003& &11.64& &0.0\\
\hline
$\delta=10^{-4}$& & & & & & & & &\\
\hline
${\mathcal{B}_p}/{10^{-9}}$& 1.6& &0.7& &5.8& &0.025& &0.0\\
${\mathcal{B}_k}/{10^{-2}}$& 15.5& &2.9& &13.5& &0.04& &0.0\\
\end{tabular}
\end{center}

\label{table1}
\end{table}

In particular we shall consider values in the region $\mathcal{B}_{p}\approx 10^{-10}$ for the weak drag due to the interaction with lattice phonons and $\mathcal{B}_{k}\approx 10^{-3}$ for the strong drag due to Kelvin wave excitation.
For the core, on the other hand we shall consider electron scattering off vortex cores as the main source of dissipation \citep{ALS}. In this case we take for the mutual friction coefficient \citep{trev}:
\be
\mathcal{B}_c=4\times 10^{-4}\left(\frac{\delta m_\p^{*}}{m_\p}\right)^2\left(\frac{\delta m^{*}_\p}{m_\p^{*}}\right)^{1/2}\left(\frac{x_\p}{0.05}\right)^{7/6}\rho_{14}^{1/6}
\ee
where $m_\p^{*}$ is the effective proton mass, $\delta m_\p^{*}=m_\p^{*}-m_\p$ and $\rho_{14}$ is the density in units of $10^{14}$ g/cm$^3$.
However, given that we are already taking the drag parameter to be a constant in the crust we shall also consider constant drag in the core. This choice is more than adequate given that we are not using a realistic prescription for the density profile or the effective mass. We shall thus consider drag parameters in the region of $\mathcal{B}_c\approx\mathcal{R}_c\approx 5\times 10^{-4}$.
Note that microscopically the exact nature of the superconductor in the neutron star interior is still open to debate \citep{PBJ06}. The values we use are consistent with the core being in a type I superconducting state \citep{SedI} (although see \citep{PBJ06} for a discussion of strong drag in a type I superconductor) and with the possibility that one has a type II superconductor and the drag is strong, either as a consequence of the magnetic field geometry \citep{trevnew} or due to a significant number of vortices cutting though the magnetic flux tubes. We do not consider here the possibility that vortices are strongly pinned to flux tubes in a type II superconductor and the most of the core is thus decoupled from the crust. This is clearly an interesting possibility which shall be the object of future work.

In our model the exact value of the drag parameter acting on a vortex section depends critically on the region we are considering, as not only do the parameters depend on density, but the very nature of the drag varies from crust to core. We shall therefore integrate over a whole vortex and define an "effective" drag coefficient which depends only on cylindrical radius (as shown in figure \ref{lag2}) by averaging over the vortex length. Our effective drag parameter thus depends mainly on how sizable a portion of the vortex is immersed in the core (i.e. the stronger drag region) rather than in the crust.

\section{Description of the model}

\subsection{Core}
Let us now describe the evolution of the system as it spins down due to the electromagnetic torque. 
First of all let us focus on the central regions of the star. Here the vortices will stretch through the fluid core and only a small portion will experience pinning to the crustal lattice. This setup is in fact very similar to that used to study the spin-up of superfluid helium in a container. In this case the vortices are only pinned to the surface of the container and unpin once the Magnus force integrated over a vortex is approximately equal in magnitude to the tension (see e.g. \citet{Glaberson, Sonin}):
\be
T_L\approx\frac{\rho_n\kappa^2}{4\pi}\ln\left(\frac{b}{\xi}\right)
\ee
with $b$ the intervortex spacing and $\xi$ the vortex core radius. This is in fact natural as we can expect that, once the Magnus force exceeds the tension the vortex will no longer behave rigidly but we can expect reconnection and Kelvin waves to be excited, possibly developing turbulence and leading to a number of vortices to unpin and begin to "creep" radially outwards \citep{Schwarz} (although bundles of vortices could be significantly more rigid \citep{RudV}).
For average parameters the Magnus force in the core will exceed the line tension almost immediately once a lag begins to develop ($\Delta\Omega_c\approx 10^{-13}$), leading to the conclusion that (as in $^4$He) unpinned vorticity dominates for most of the time, and re-pinning can only take place if the the two components are essentially comoving (i.e. $\Delta\Omega<\Delta\Omega_c$).
Note the the situation is radically different in the crust, where the vortex line is fully immersed in the lattice and is pinned in several points. In this case we can expect it to move freely only once the maximum pinning force of section \ref{pin}  is exceeded.


Let us, however, stress that the assumption of free vortices is not crucial for our model.
One can, in fact, easily account for the fact that possibly not all vortices will be free to move, but at any given time only a small fraction will be free, and subject to the drag force. Following \citet{miri2,miri} we shall assume that the instantaneous number density of free vortices $n_m$ is given
\be
n_m=\xi n_v
\ee
where $n_v$ is the total number density of vortices, and $\xi$ is the fraction of unpinned vortices at a given time.  By now averaging over time we can obtain an effective drag parameter $\mathcal{B}_E=\xi\mathcal{B}$. 

For the standard "creep" model \citep{Alpar1,LinkEp} the unpinning probability takes the form
\be
\xi\approx \exp\left(-\frac{E_p}{kT}(1-\frac{\omega}{\omega_c})\right)
\ee
where $E_p$ is the Energy barrier that the vortex needs to overcome, and $T$ is temperature, $\omega$ is the lag between the components, and $\omega_c$ is the critical lag. Unfortunately there are huge theoretical uncertainties on the unpinning probability and observational constraints are also very model dependent. However by varying $\xi$ and thus reducing the 'effective' drag parameter we can account for vortex ``creep'' in the linear regime, which has been used to interpret the relaxation phase of Vela glitches \citep{Alp89}.

The details of the steady state are, however, not crucial for our formulation, which is fortunate as not only vortex creep but also vortex repinning is still poorly understood, although efforts are being made to tackle this problem \citep{Sedrakian,Linkn}.
Eventually the system will settle down to an equilibrium state in which both fluids are spinning down at the same rate and the lag can be estimated, from the equations in (\ref{eom1}) and (\ref{eom2}), to be
\be
\Delta\Omega(\tilde{r})\approx \frac{\dot{\Omega}}{2\mathcal{B}(\tilde{r})\Omega}
\label{lag}\ee
where $\mathcal{B}$ is the effective (averaged) mutual friction parameter and we have neglected the effect of entrainment and differential rotation. We will see in the following that this is actually a very good approximation to the equilibrium lag.

\subsection{Strong pinning region}

The repinning region, i.e. the equatorial region in which the pinning force rises steeply, deserves a separate discussion. As the vortices encounter what is, effectively, a barrier, we shall assume that they repin almost immediately. As we discuss below this is effectively a situation in which the vortices 'creep' steadily outwards, i.e. move outwards with, on average, a low velocity. This suggests that Kelvin-phonon interactions will be strongly suppressed and that we can consider weak mutual friction coefficients, due to interactions with lattice phonons, in the region of $\mathcal{B}_p\approx\mathcal{R}_p\approx10^{-10}$, as described in the previous sections.

In fact, as the drag description is not valid close to repinning \citep{Linkn}, one could argue that the form of the mutual friction in (\ref{mf}) is also not valid. Strictly speaking this is true but, as the full problem of the motion of a vortex in a strong pinning potential is at present intractable, we shall assume that the coupling given by mutual friction with the weak drag given by the interaction with phonons in the lattice is a good approximation.
The whole discussion could also be made rigorous by following the approach of \citet{Lagrange} and assuming that an extra 'pinning' force is acting on the vortices. This leads to an effective mutual friction parameter
\be
\mathcal{B} = { 1 \over 1 + \mathcal{R}^2 } \left[ \mathcal{R} + { a_\parallel + a_\perp \mathcal{R} \over w } \right]
\ee
which now depends on the relative velocity $w$ and the components of the effective force that describes the motion through the pinning potential, $a_\parallel$ (along the relative flow) and $a_\perp$ (perpendicular to the relative flow). Note that this is not the actual "pinning" force, but rather an effective force that mimics the effect of creep.
The vortex velocities then take the form
\be
v_\parallel\equiv a_\parallel \mathcal{R} -  w - a_\perp
\ee
and
\be
v_\perp\equiv a_\parallel +  \left( w + a_\perp  \right) \mathcal{R}
\ee
and in particular one finds that
\be
\mathcal{B}\approx \frac{v_\perp}{w}
\ee

However, given that we do not have a consistent description of an effective pinning force (but only the maximum pinning force) we shall simply assume that all vortices are pinned below the critical velocity, and then relax to their steady state configuration once the critical unpinning lag is exceeded. This is equivalent to assuming that below the critical velocity only a very small fraction of the vortices can 'creep', which is reasonable due to the sharp rise in the pinning force.

This leads to the neutrons relaxing to their steady state rotational rate on a timescale $\tau_r\approx 1/2\Omega \mathcal{B} \approx 10^8$ s (for $\mathcal{B}\approx 10^{-10}$), during which the protons have spun down and the unpinning "front" has moved out along the strong pinning equatorial region. Essentially the unpinning front moves radially out at a speed $v_r=\Delta r/\tau_g$ with $\tau_g=\Delta \Omega_{max}/\dot{\Omega}$, which leads to a speed of approximately $v_r\approx \dot{\Omega} \Delta r / \Delta\Omega_{max} \approx 10^{-4}$ cm/s, where $\Delta r=400$m is the width of the strong pinning region, and $\Delta\Omega_{max}\approx 10^{-2}$ is the maximum in the lag in the equatorial region.
This means that the vortices that have unpinned from parts of the star closer to the rotation axis are clustered in this region, over a lengthscale $l\approx \tau_r v_r \approx 10^4$ cm. We shall assume that it is this region that gives rise to the glitch once the maximum lag is exceeded and the vortices can no longer be pinned, thus no longer 'creep', but are free to escape radially outwards.

\subsection{Glitch}

\begin{figure}
\includegraphics[height=7cm,clip]{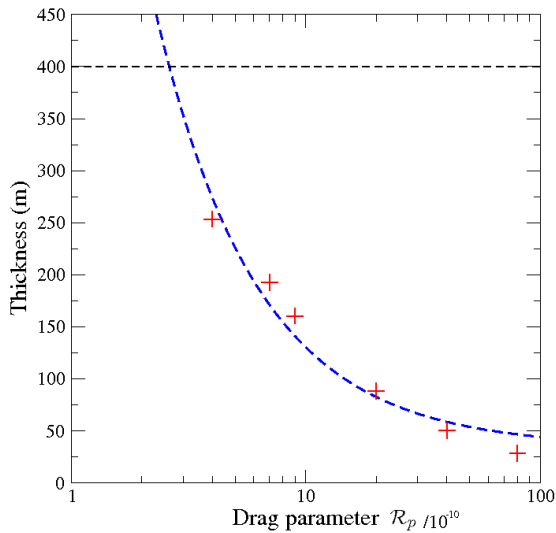}
\caption{We plot the extent of the region in which Kelvin oscillations can be excited as a function of the weak drag parameter $\mathcal{R}_{p}$. Note that the whole strong pinning region is 400 meters thick, as indicated by the horizontal line. We extrapolate the value of the drag parameter for which the whole region is involved by fitting a function of the form, inspired by the estimate in (\ref{estim}), $f(x)=a+b/(\mathcal{R}_{p}/10^{-10})$, where the best fit parameters were $a=34.77\pm 13.7$ m and $b=955.8\pm 101.9$ m. As we can see the whole strong pinning region participates in the glitch for $\mathcal{R}_{p}<2\times 10^{-10}$, as expected from the analytical estimates.}
\label{fit}
\end{figure}

Once the maximum lag has been reached we shall assume that the pinning force can no longer contrast the Magnus force and that the vortices move out. They very rapidly reach a state of co-rotation with the neutrons, leading to velocities of $\approx 10^4$ cm/s with respect to the lattice. In this regime Kelvin-phonon dissipation will certainly dominate and recouple the neutrons and protons on a very short timescale, giving rise to the rapid exchange of angular momentum that characterises the glitch.
Consistently with our prescription for the electron-phonon dissipation we shall take the mutual friction coefficient for Kelvin-phonons to be $\mathcal{B}_k\approx10^{-3}$, and assume that acts only in the region that has not relaxed to a steady state, i.e. the region in which the lag is greater than the equilibrium value:
\be
\Delta\Omega>\frac{\dot{\Omega}}{2\mathcal{B}_{p}\Omega}
\label{estim}\ee
We can estimate the relaxation timescale in the crust as $\tau=1/2\Omega\mathcal{B}_{p}$ and assume that the pinning front moves through the strong pinning region over a period of 3 years, i.e. with a velocity $v\approx 4\times 10^{-4}$ cm/s. The region that has not yet relaxed will then have a thickness of 
\be
\Delta\approx v\tau\approx\frac{3\times 10^{-6}}{\mathcal{B}_{p}}\label{delta}
.\ee
For a drag parameter $\mathcal{B}_p\approx 10^{-10}$ one would then have approximately all of the strong pinning region involved in the glitch. A more quantitative analysis is shown in figure \ref{fit} where we show the edge of the non relaxed region as a function of the drag parameter, obtained by measuring where the lag between neutrons and protons deviates from the analytic estimate in (\ref{estim}). We cannot extend the numerical analysis to very weak drag, nevertheless we see that the region becomes larger for weaker drag and that extrapolating our results we would have the whole crust involved in the glitch for $\mathcal{B}_p\approx 2\times 10^{-10}$. The extrapolation was obtained by fitting a function of the form, inspired by the estimate in (\ref{estim}), $f(x)=a-b/\mathcal{B}_p$, where the best fit parameters were $a=2330.45\pm 26.14$ m and $b=1911.67\pm 203.9$ m, which are in reasonable agreement (within factors of a few) with the analytical estimate. We shall thus assume that for $\mathcal{B}_p<2\times 10^{-10}$ Kelvin oscillations of the vortices can be excited in the whole strong pinning region.

\subsection{Post-glitch relaxation}

Subsequent to the glitch neutrons and protons will still be in approximate co-rotation only in the interior regions that are coupled on timescales shorter than the glitch rise time. In the exterior core the lag will have decreased by a factor $\Delta\Omega\approx 10^{-4}$ due to the glitch itself. The neutrons and protons then return to their steady state configurations with the timescale appropriate to the mutual friction parameter $\mathcal{B}$ at the cylindrical radius we are considering, thus naturally recoupling the core on timescales that range from the glitch rise time in the innermost regions to months in the outermost regions (i.e. $\tilde{r}\approx 11$ km). 
As far as the strong pinning region is concerned we shall assume that the vortices repin after the glitch, and gradually unpin (or begin to 'creep') again as a lag builds up due to the protons spinning down. As already discussed the details of the repinning are highly uncertain, but given the weak coupling in the crust the details of such a mechanism will only impact on the longer relaxation timescales.

\section{Results}

In order to set up the pre-glitch conditions for our system we impose that the two fluids rotate at the same rate and then begin to evolve in time. This would clearly not be the correct condition if, say, we are considering the situation immediately following a previous glitch. However, as we are setting up the situation for the next glitch we are only interested in obtaining a steady state on long timescales (i.e. the inter-glitch timescale, which is approximately 3 years for the Vela pulsar). As we intend to evolve the system for approximately 3 years we cannot evolve the whole interior, which would require us to deal with coupling timescales of seconds, but assume that it is always coupled to the crust by including its moment of inertia in that of the charged component, and only evolve the region between $\tilde{r}=11.14$ Km and $\tilde{r}=11.55$ Km (i.e. the outer core in which the value of the drag parameter becomes small and varies rapidly, and the strong pinning region of the inner crust). We neglect the outer crust, which we do not expect to be a bad approximation as its moment of inertia is small compared to that of the other regions of the star. The equations of motion are thus integrated only up to the maximum of the pinning force, where we impose that the spatial derivative of the neutron angular velocity must vanish, i.e. $\frac{\partial{\Omega_\mathrm{n}}}{\partial{r}}=0$, which is appropriate if the vortices are still pinned at that point. At every time-step we thus evaluate the integral in (\ref{integrate1}) and then evolve the coupled equations in (\ref{integrate1}-\ref{integrate}) forward in time with a four step Runge-Kutta algorithm.  
 
\begin{figure}
\includegraphics[height=7cm,clip]{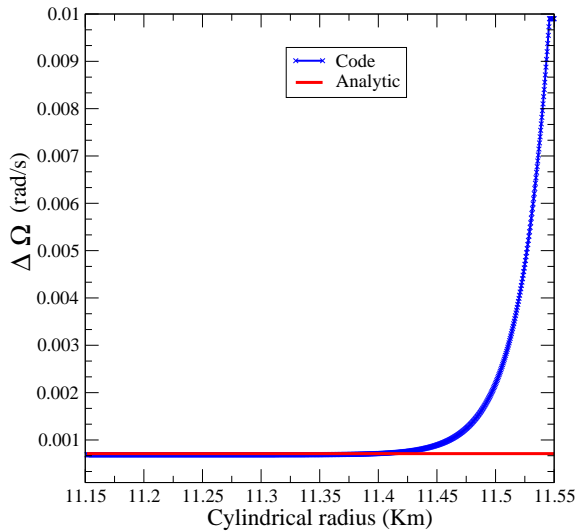}
\caption{We compare, in the strong pinning region of the crust, the analytical prediction for the lag $\Delta\Omega$ between protons and neutrons to the value obtained in our simulations just before the glitch. The calculation was performed for $\mathcal{B}_\p=10^{-9}$. As can be seen from the figure, the analytical estimate is a very good approximation, except for the outer region in which the system has not yet reached equilibrium. It will thus be a good approximation for the core, in which the coupling timescale between neutrons and protons is considerably shorter than in the crust.}
\label{lagfig1}
\end{figure}

Once the maximum of the lag in the strong pinning region is reached we initiate the glitch by switching to strong Kelvon drag ($\mathcal{B}_k$) in the region that has not relaxed, i.e. the region for which $\Delta\Omega>\frac{\dot{\Omega}}{2\mathcal{B}_{p}\Omega}$. At this point we track the evolution of the whole core by extending the numerical grid and filling the region that had not evolved with the prescription $\Delta\Omega=\frac{\dot{\Omega}}{2\mathcal{B}_c\Omega}$. As can be seen in figure (\ref{lagfig1}) this is indeed a good approximation for the initial condition, given that the timescale on which the interior reaches a steady state is considerably shorter than the inter-glitch timescale.
When the strong drag regions have reached their equilibrium lag we switch off the Kelvon drag and assume that the vortices have repinned. We then follow the immediate post-glitch relaxation by evolving the whole star over a timescale of days. 

To summarise, we set up a system of co-rotating fluids with pinned vortices in the crust and evolve it in time. As the lag increases the depinning front moves thorugh the crust. Once the maximum critical lag is reached the vortices move rapidly outwards and strong Kelvon mutual friction recouples the components, giving rise to the glitch. After the glitch the lag has increased throughout the star and each region relaxes to equilibrium on a timescale determined by the local values of the (average) mutual friction.

\subsection{The Vela Pulsar}

\begin{figure}
\centerline{\includegraphics[height=7cm,clip]{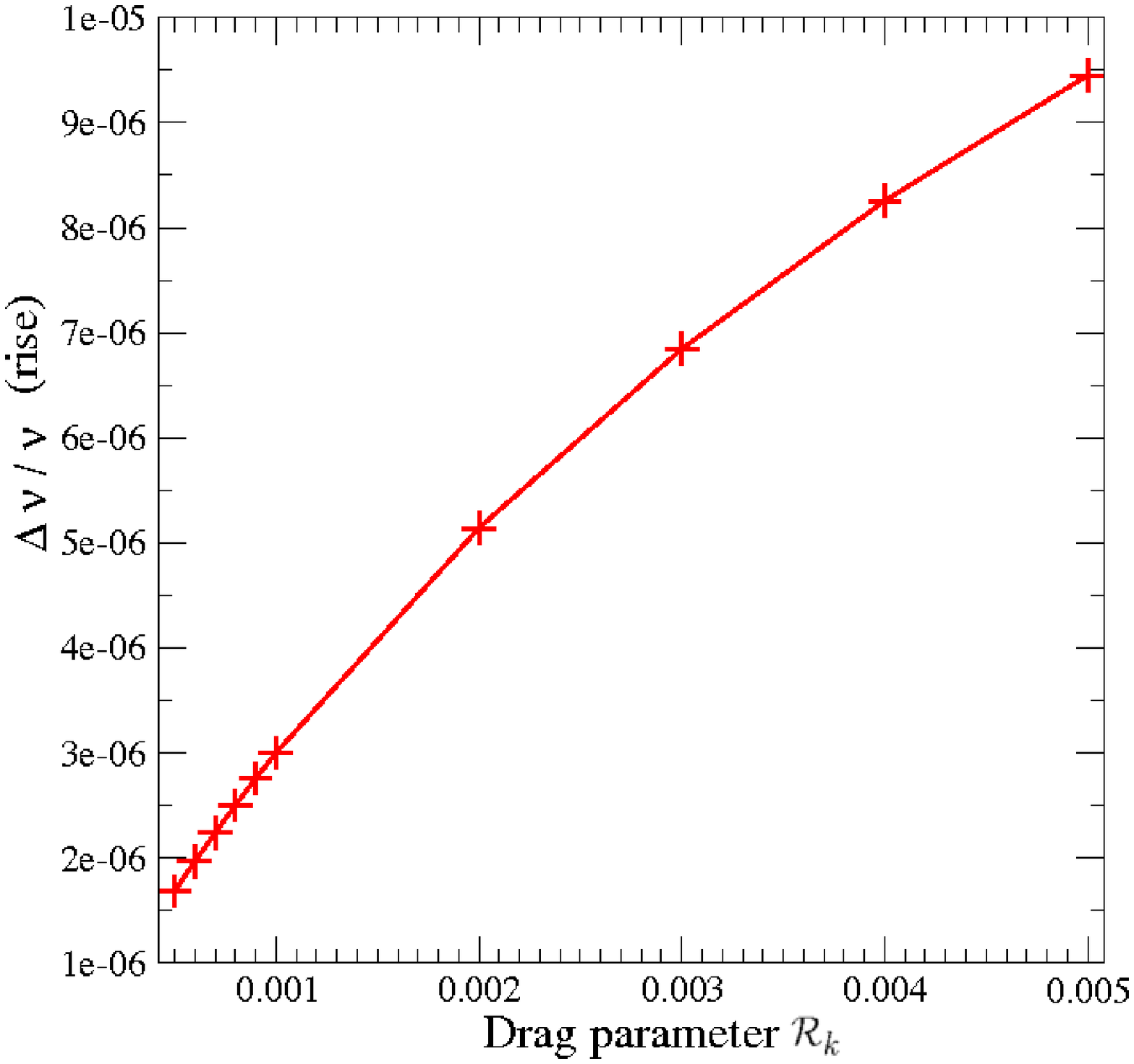}}
\centerline{\includegraphics[height=7cm,clip]{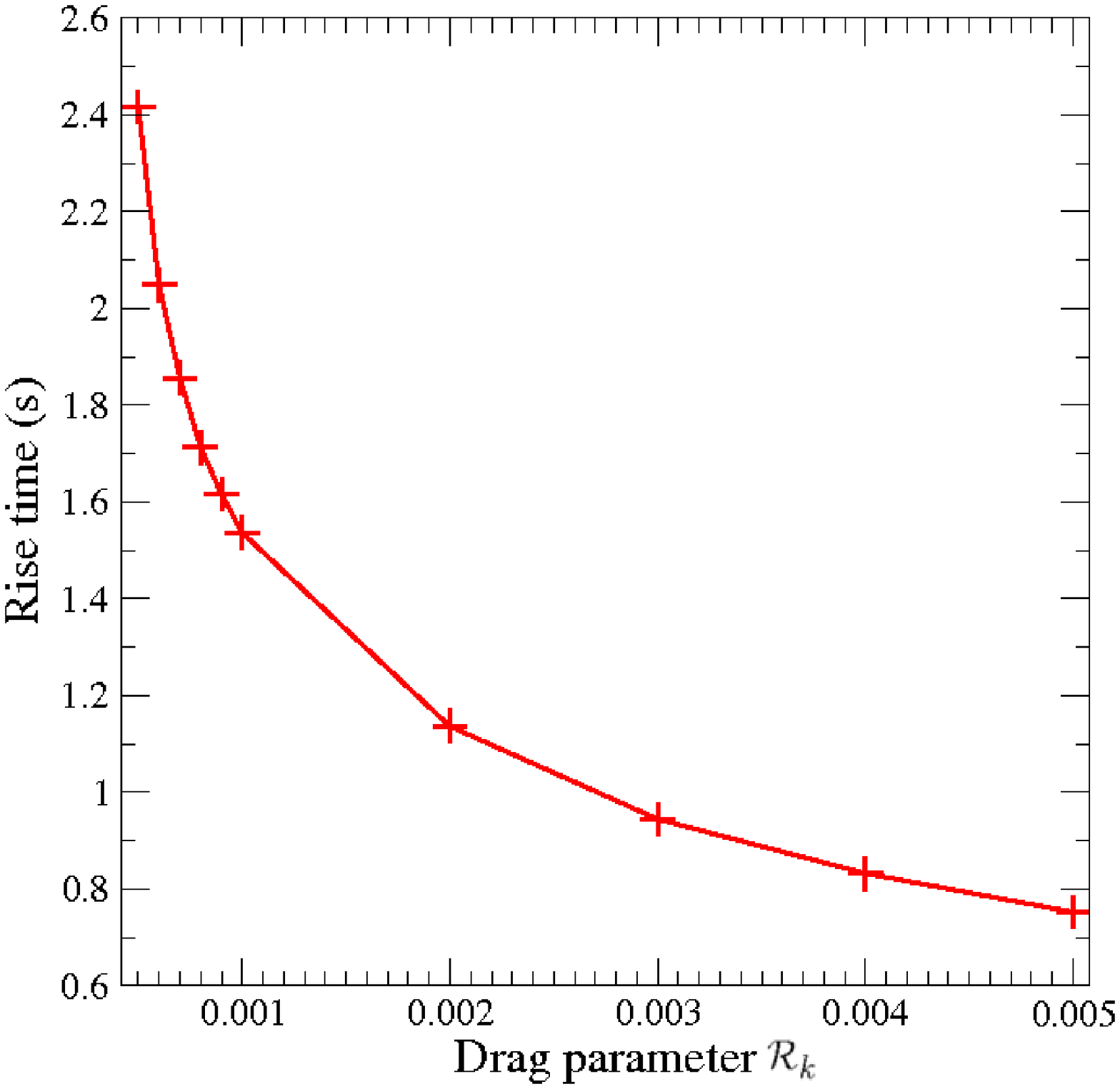}}
\caption{In the top panel we plot the size of the glitch (at the end of the rise) and in the bottom panel the rise time, both as a function of the strong Kelvin drag parameter $\mathcal{R}_{k}$. For these simulations we have take $\mathcal{R}_{p}=10^{-10}$ and $\mathcal{R}_{c}=5\times 10^{-4}$.}\label{inst}
\end{figure}

Let us first of all examine the case of the Vela Pulsar, which is the prototype system for giant glitches. The Vela (PSR B0833-45 or PSR J0835-4510) has a spin frequency $\nu\approx 11.19$ Hz and spin-down rate $\dot{\nu}\approx -1.55 \times 10^{-11}$ Hz s$^{-1}$. Giant glitches are observed roughly every two to three years, and have relative frequency jumps of the order $\Delta \nu/\nu\approx 10^{-6}$. The spin-up is instantaneous to the accuracy of the data, with the best constraint being an upper limit of 40s for the rise time obtained from the 2000 glitch \citep{dodson} (a similar upper limit of 30 s was obtained from the analysis of the 2004 glitch, but was less significant due to the quality of the data \citep{dodson2}). The glitch is usually fitted to a model consisting of permanent steps in the frequency and frequency derivative and a series of transient terms that decay exponentially. It is well known that at least three are required to fit the data, with decay timescales that range from months to hours \citep{Flanagan}. Recent observations of the 2000 and 2004 glitch have shown that an additional term is required on short timescales, with a decay time of approximately a minute. The fitted values for the relaxation of these two glitches are shown in table \ref{glitches}. The spindown rate always increases after a glitch with larger relative increases (up to a factor of 10) which decay on the shorter timescales and smaller (a few percent) increases that decay on the longer timescales of days and months. Note that a further two glitches were detected in 2008 and 2010, but no timing analysis has yet been published.
\begin{table}
\caption{ The fitted values for the relaxation of the Vela 2000 and Vela 2004 glitches, from \citet{dodson} and \citet{dodson2}. After removing the pre-glitch spindown the fit on the residuals is performed with a function of the type $f(t)=\Delta_pF+\Delta_p\dot{F} t +\sum_i f_i \exp{(-t/\tau_i)}$.}
\begin{center}
\begin{tabular}{l l l l}
\hline
 &2000& &2004\\
\hline
$\Delta_pF$/Hz &3.45435E-05& &2.2865E-05\\
$\Delta_p\dot{F}$/Hz s$^{-1}$& -1.0482E-13& &-1.0326E-13\\
$f_1/10^{-6}$Hz&0.02& &54\\
$f_2/10^{-6}$Hz&0.31& &0.21\\
$f_3/10^{-6}$Hz&0.193& &0.13\\
$f_4/10^{-6}$Hz&0.2362& &0.16\\
$\tau_1$&1.2$\pm 0.2$ mins& &1$\pm 0.2$ mins\\ 
$\tau_2$&0.53 days& &0.23 days\\
$\tau_3$&3.29 days& &2.10 days\\
$\tau_4$&19.07 days& & 26.14 days
\end{tabular}
\end{center}

\label{glitches}
\end{table}

\begin{figure}
\includegraphics[height=7cm,clip]{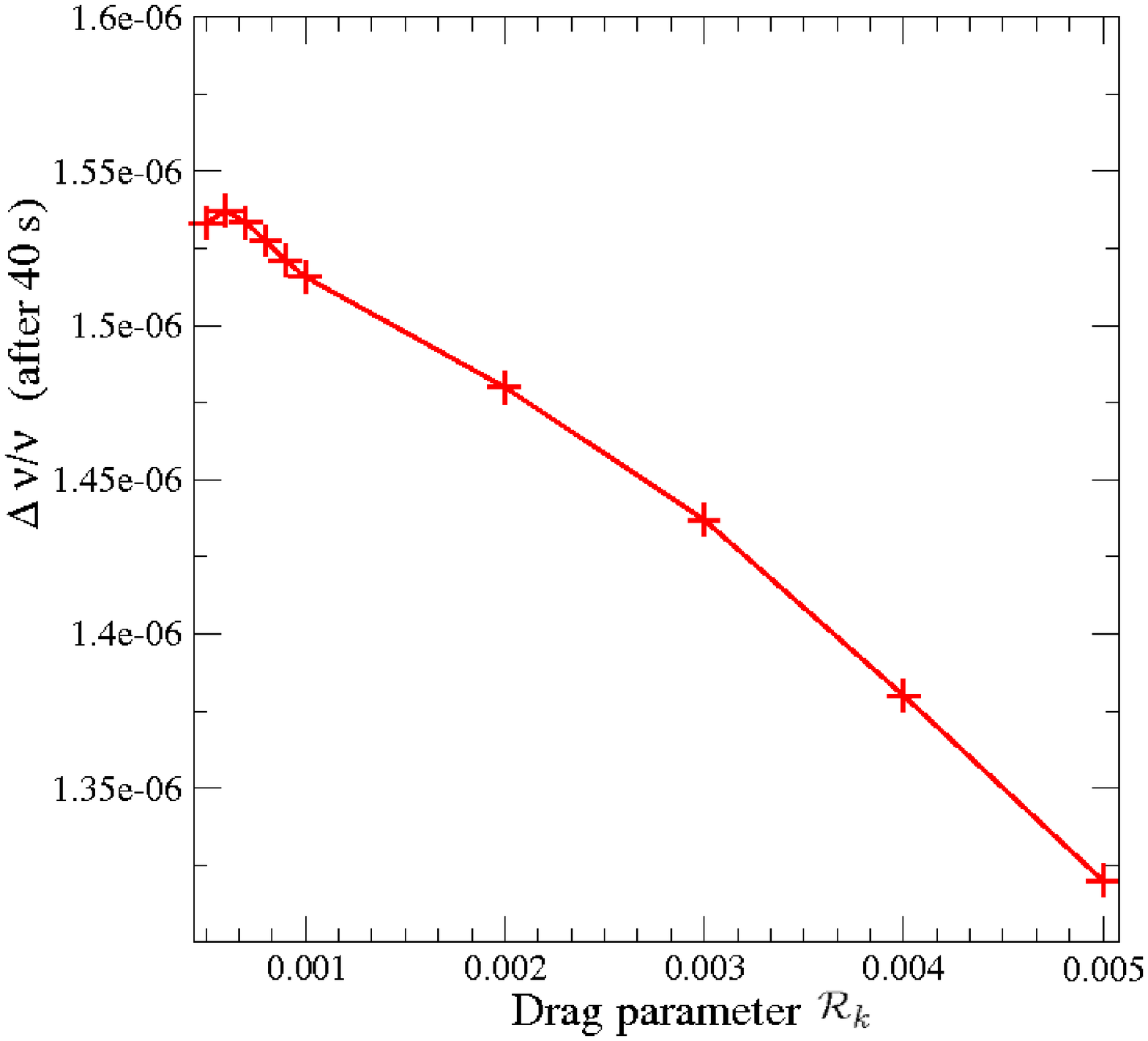}
\includegraphics[height=7cm,clip]{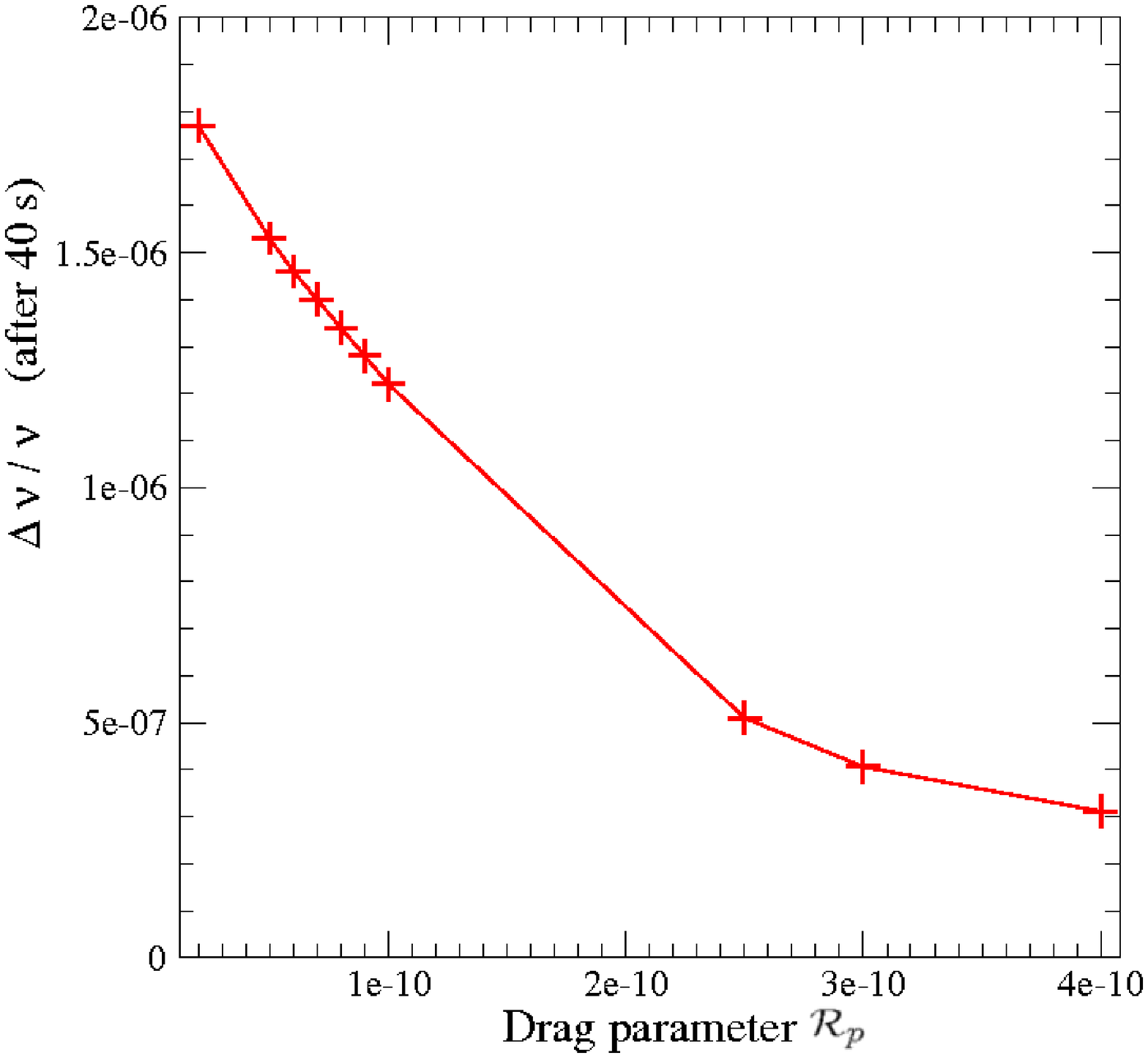}
\caption{The size of the glitch, extracted 40 s after it is initiated, as a function of the drag parameters. In the top panel we plot the glitch size as a function of the strong drag parameter $\mathcal{R}_k$. In this case the stronger the drag the smaller the glitch, as an initially larger glitch relaxes faster and is, in fact, smaller after 40 s. The remaining drag parameters were taken to be $\mathcal{R}_p=5\times 10^{-11}$ and $\mathcal{R}_c=4\times 10^{-4}$. In the bottom panel we show the glitch size as a function of the weak drag parameter $\mathcal{R}_p$. Again for larger values of the drag we obtain a smaller glitch, as in this case more of the fluid in the crust has relaxed to its equilibrium configuration prior to the glitch and less angular momentum is available to be exchanged. The remaining drag parameters were taken to be $\mathcal{R}_k=6\times 10^{-4}$ and $\mathcal{R}_c=4\times 10^{-4}$.}\label{40s}
\end{figure}

\begin{figure}
\includegraphics[height=7cm,clip]{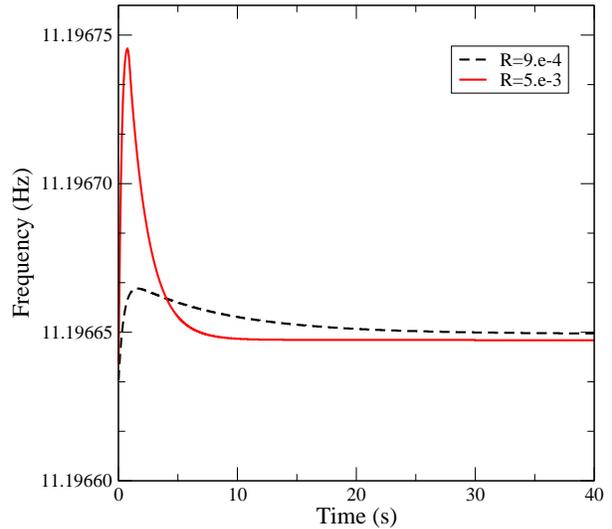}
\caption{The first 40 s of the glitch for two values of the Kelvin drag parameter $\mathcal{R}_k$. A stronger drag (i.e. a shorter coupling timescale) gives rise to an initially larger glitch that however rapidly relaxes to a lower spin rate than that of a glitch involving a weaker drag parameter.}
\label{risef}
\end{figure}

\begin{figure}
\includegraphics[height=7cm,clip]{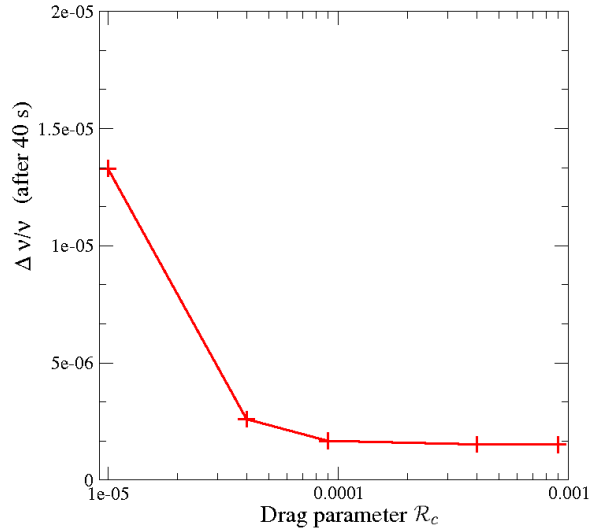}
\caption{The size of the glitch, extracted 40 s after it is initiated, as a function of the core drag parameter. Weaker drag parameters give rise to larger glitches as less of the core can contribute to the moment of inertia during a glitch. The remaining drag parameters were set at $\mathcal{R}_p=5\times 10^{-11}$ and $\mathcal{R}_k=1\times 10^{-3}$.}\label{40s2}
\end{figure}

Consistently with the results of \citet{pinning1}, \citet{pinning1phd} and \citep{Pierre1} we take the maximum of the critical unpinning lag to be $\Delta\Omega_c=10^{-2}$, which naturally leads to a glitch recurrence time of roughly 3 years (the exact time depends on the choice of drag parameters, but is always approximately 1100 days). Our simulations can then reproduce glitches with fractional rises $\Delta\nu/\nu\approx 10^{-6}$ for a range of parameters. In figure \ref{inst} we show the rise time and the 'instantaneous' size of the glitch, i.e the size at the end of the rise, as a function of the strong Kelvin drag parameter $\mathcal{R}_\mathrm{strong}$. As expected the larger the drag parameter the shorter the rise time and the larger the glitch. This is simply due to the fact that the faster the glitch (or rather the stronger the Kelvin drag parameter with respect to the drag in the core) the smaller the fraction of the core superfluid neutrons that can remain coupled to the crust and contribute to its moment of inertia during the glitch. 

\begin{figure}
\includegraphics[height=7cm,clip]{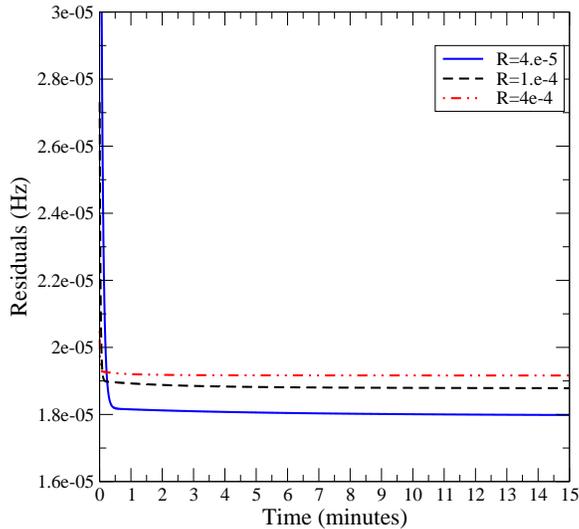}
\caption{We plot the post-glitch residuals after subtracting the pre-glitch spindown. The top panel shows the first 15 minutes of the relaxation for three values of the core drag parameter $\mathcal{R}_\mathrm{core}$. We can see that a weaker drag in the core leads to less of the superfluid coupling to the crust and thus to a larger glitch, that however relaxes, on a timescale of several minutes, more than in the case of stronger coupling.}
\label{relax1}
\end{figure}
\begin{figure}
\includegraphics[height=7cm,clip]{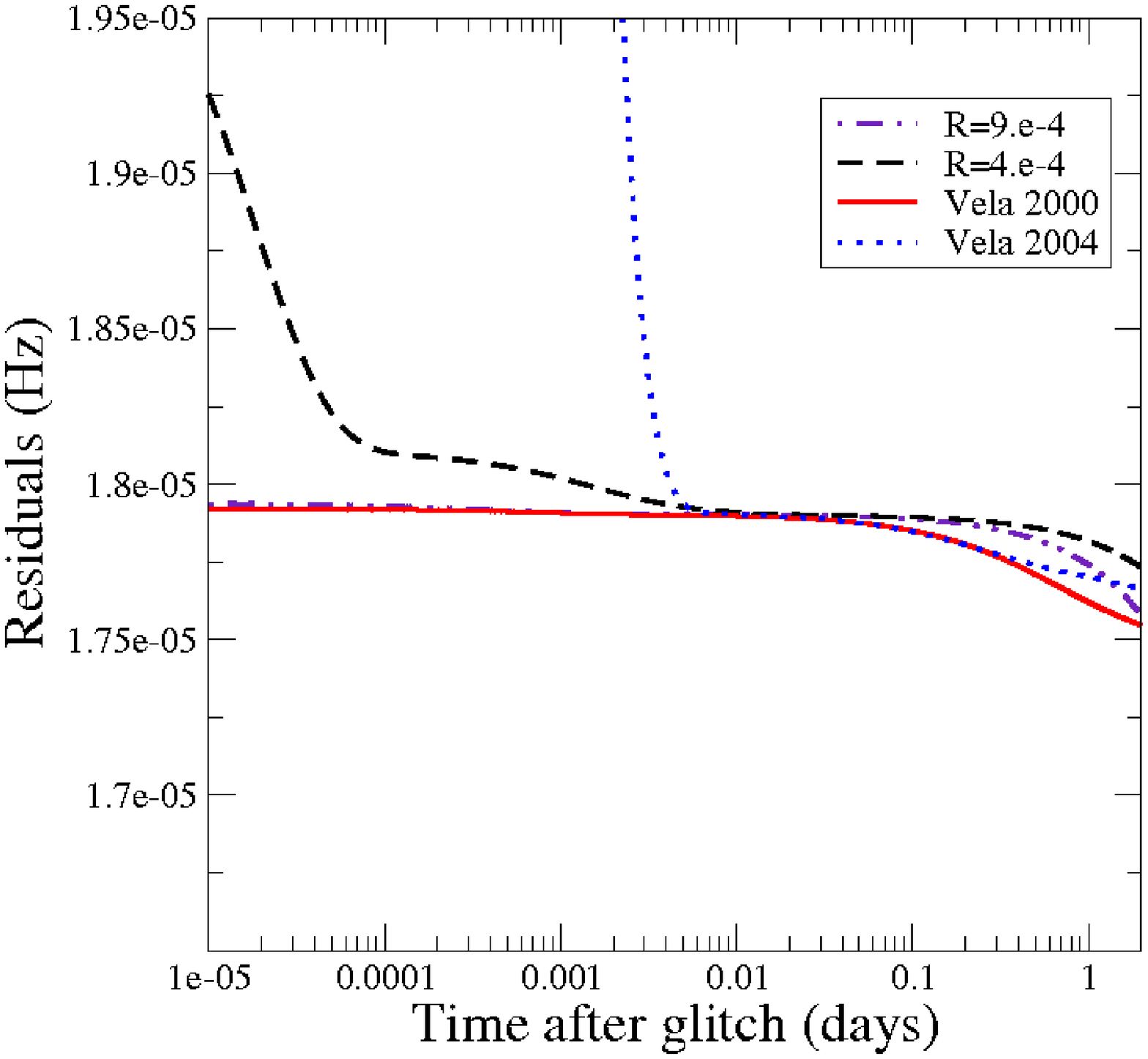}
\caption{The post-glitch residuals, rescaled as in figure \ref{compA}, but for a longer timescale of 2 days.}
\label{long}
\end{figure}

However a strong Kelvin drag will lead to rise times of the order of seconds, well below the observational upper limit of 40 s and thus well below the observational capabilities of current radio telescopes that can, for the Vela pulsar, at best deal with 10 s folds. The rise will then be followed by a rapid relaxation. In fact if we then extract the glitch size 40 s  after the glitch is initiated, as shown in figure \ref{40s}, we see that the situation is reversed. The stronger the drag the smaller the glitch after 40 s, as an initially larger jump in frequency relaxes faster and leads, in fact, to a lower frequency after 40 s, which is illustrated in figure \ref{risef}. In figure \ref{40s} we also see that stronger phonon drag parameters in the core gives rise to smaller glitches, as more of the crust has relaxed to its equilibrium configuration and less angular momentum is available to be exchanged.

\begin{figure}
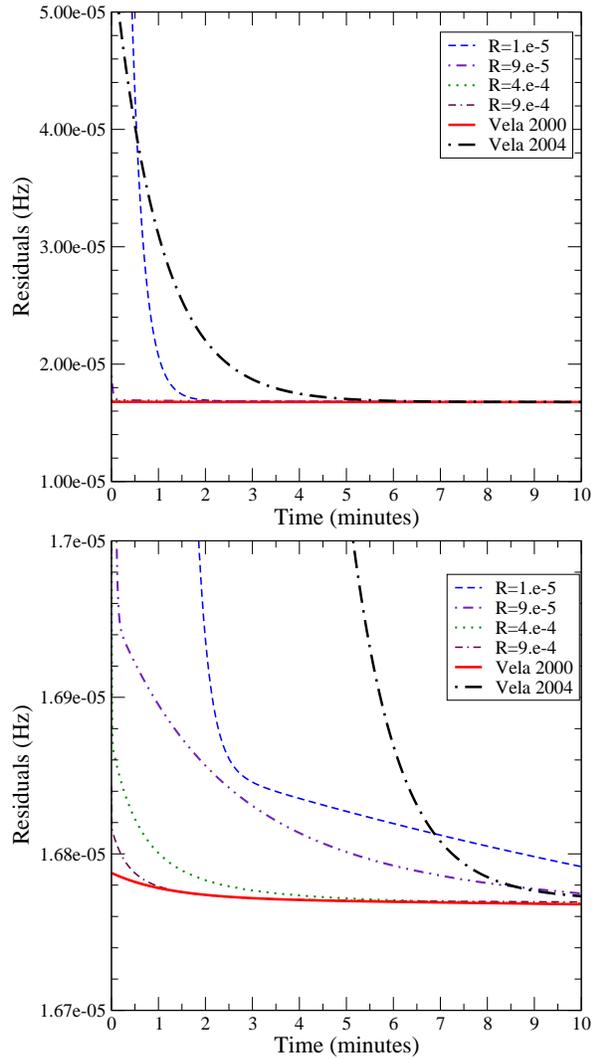

\includegraphics[height=7cm,clip]{tscale1.eps}
\includegraphics[height=7cm,clip]{tscale2.eps}
\caption{We plot the post-glitch residuals after subtracting the pre-glitch spin-down and having scaled all the results so that the glitch has decayed to the same frequency after 10 minutes. The bottom panel is a zoom in of the top panel. It is clear that stronger drag parameters provide a better fit to the Vela 2000 relaxation fit (for which the data was of better quality) and that very weak drag parameters are still excluded by the Vela 2004 relaxation fit. It would appear that the observations are consistent with a drag parameter in the range $\mathrm{R}\approx 10^{-4}-10^{-3}$, as expected theoretically for electron scattering off vortex cores. A very weak drag, due to only a small fraction of the vortices 'creeping', would not appear to be consistent with the short timescale components of the relaxation.}\label{compA}
\end{figure}

\begin{figure}
\includegraphics[height=7cm,clip]{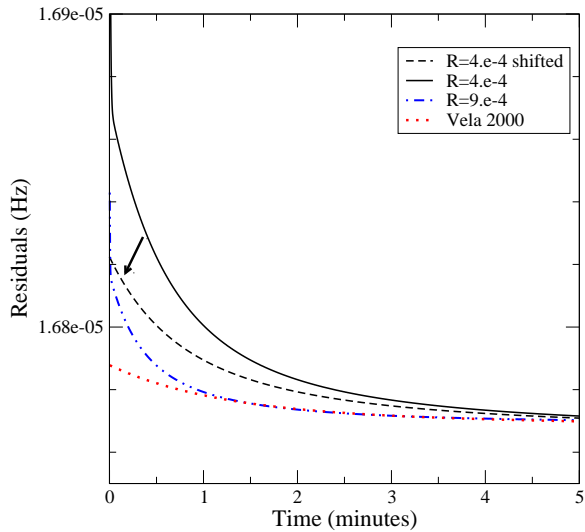}
\caption{We plot the post-glitch residuals, rescaled as in figure \ref{compA}, to illustrate the degeneracy between the assumed glitch epoch and the strength of the drag in the core. We show the effect of assuming that the glitch has occurred 30 s before the assumed epoch. We can see that the curve for $\mathcal{R}_c=4\times 10^{-4}$ now produces a smaller glitch and gives a better fit to the Vela 2000 data. The curve thus appears closer to that given by taking $\mathcal{R}_c=9\times 10^{-4}$ but not adjusting the glitch epoch.}
\label{shift}
\end{figure}

While the magnitude of the strong Kelvin drag affects mainly the glitch size and rise time, the strength of the drag in the core will affect, as shown in figure \ref{40s2} and \ref{relax1} the size of the glitch itself (as it determines the amount of core superfluid that participates in the glitch) but also crucially affect the short term post glitch relaxation. 
In figure \ref{relax1} we can see that a weaker drag in the core leads to less of the superfluid coupling to the crust and thus to a larger glitch, that however relaxes, on a timescale of several minutes, more than in the case of stronger coupling. Unfortunately the short timescale component of the relaxation is not strongly constrained by observations, as it was only measurable in the 2000 and 2004 glitches, with vastly different magnitudes, as can be seen from  figures \ref{compA} and \ref{long}. The results in figure \ref{compA} would however indicate that a core drag parameter $\mathcal{R}\approx 10^{-5}$ is still marginally consistent with observation, but that there is no evidence for a much weaker drag, due for example to the fact that most vortices are pinned (either to the crust or to flux tubes in the core) and only a small fraction of them can creep. It would thus appear that the observations of the short term relaxation are consistent with a mutual friction drag in the range $\mathrm{R}\approx 10^{-4}-10^{-3}$, consistent with theoretical expectations for electron scattering off vortex cores. This is encouraging, given the approximate treatment of the drag parameter in this work. Note, however, that quantitative statements are not easy to make as there is a large degeneracy between the various parameters that enter the model. For example in figure \ref{shift} we show that the uncertainty on the time of the glitch, which is of about 30 s, can lead to an uncertainty on the value of the drag parameters.

In figure \ref{long} we show the longer timescale components of the relaxation. These also appear consistent with a strong core drag due to electron scattering, although a slower spindown may be preferred. Quantitative statements regarding the longer relaxation timescales (days to months) are, however, hindered by computational and theoretical issues. On the one hand evolving the full system of coupled equations for longer than a few days is computationally challenging, on the other theoretical uncertainties regarding repinning after a glitch and creep, begin to have a significant impact on the results for these longer timescales, while they do not affect short timescales of minutes or hours. In order to obtain truly quantitative results for longer timescales it would be necessary to address these theoretical issues and also to include a realistic density dependence of the drag parameters. Finally other effects are likely to play a role in the relaxation timescale, notably friction at the crust-core interface \citep{vE1}, the inclusion of which would require us to relax the rigid rotation assumption for the charged component, which is beyond the scope of this paper but will be the focus of future work.

\section{Giant Glitchers}

\begin{figure}
\includegraphics[height=7cm,width=7cm,clip]{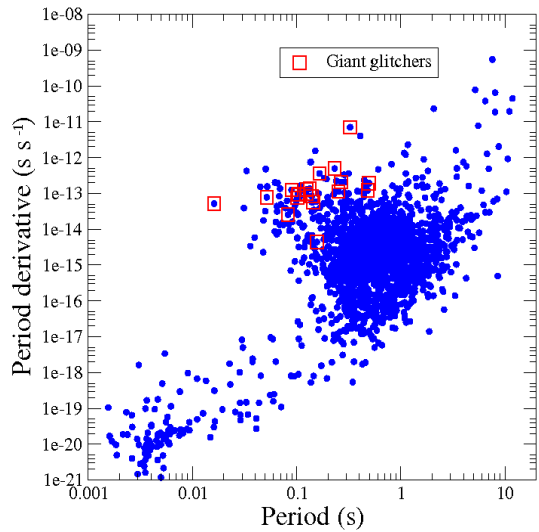}
\caption{The location of the giant glitching pulsars as identified by \citet{Espinoza} in the $P-\dot{P}$ diagram. This is the population of pulsars that show large steps in the frequency ($\Delta\nu\approx 10^{-4}$ Hz) and always exhibit an increase in the spindown rate after the glitch.}
\label{ppdot}
\end{figure}

\begin{figure}
\includegraphics[height=7cm,clip]{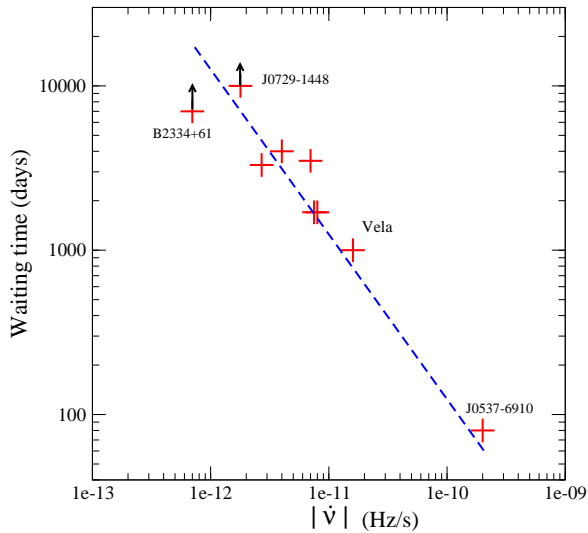}
\caption{We plot the approximate waiting time between glitches for the pulsars that have shown multiple giant glitches, as a function of the spin down rate. We also include two pulsars that have shown only one glitch but also have a long baseline for the observations and can thus provide us with an interesting lower limit on the waiting time. The data appears consistent with the notion that giant glitches can occur once a critical lag of approximately $\Delta\Omega=10^{-2}$ is reached. In fact the Vela like pulsars glitch every few years, but the X-ray pulsar J0527-6910, which is spinning down approximately an order of magnitude faster glitches every few months, while the lower limits on slower pulsars indicate that they may glitch every decade. In fact the data appears to be well described by a fit of the form y=A/x, as shown in the figure, with $y$ the waiting time in seconds, $x$ the frequency derivative and $A=1.082212\times 10^{-3}$ Hz.}\label{cristobal}
\end{figure}

So far we have only applied our model to the giant glitches of the Vela pulsar. Giant glitches have, however, been observed in several other pulsars and we would expect them to also occur when the system reaches the maximum critical lag for unpinning. As already mentioned this needs not be the case for smaller glitches that are likely to be due to random unpinning and may be, for example, described in terms of vortex avalanche dynamics \citep{NewM}. In fact recent observations support the view that there is a bimodal distribution in the glitch size, with two different populations, the "giant" glitchers and pulsars that only exhibit smaller glitches \citep{Espinoza}.
We shall thus focus on the population of ``Vela-like'' pulsars that show giant glitches, as defined by \citet{Espinoza}, which are shown on the $P-\dot{P}$ diagram in figure \ref{ppdot}. In particular 7 of these objects have multiple giant glitches, and one can thus derive an approximate waiting time between glitches. In figure \ref{cristobal} we plot the waiting time as a function of the spindown rate $\dot{\nu}$. The data appears consistent with our theoretical expectation that a pulsar that is spinning down faster will build up the critical lag on a shorter timescale and glitch more often. In particular we can see that most systems lie close to the Vela in the $P-\dot{P}$ diagram and also glitch roughly every few years, but the X-ray pulsar J0527-6910, which is spinning down approximately an order of magnitude faster glitches every few months, while the lower limits on slower pulsars indicate that they may glitch every decade.

Naturally the situation will be complicated by the fact that these pulsars also exhibit smaller glitches, which may transfer part of the angular momentum before a giant glitch, and by the fact that the critical lag will also depend, albeit weakly \citep{Stefano}, on the mass and radius of the star. 
Given these limitation our model would, however, appear to be consistent with the observed inter-(giant)glitch waiting time.

Unfortunately only the Vela is currently observed at short enough intervals to allow a fit for the short timescale transient terms in the relaxation, so more accurate tests are not currently possible for other pulsars. It would be of great interest if such observations for other giant glitchers were to become possible with the new generation of radio-telescopes such as LOFAR and the SKA.

\section{Conclusions}

We have presented a hydrodynamical two fluid model of pulsar glitches that can consistently model all phases of the glitch itself. Our model can be successfully applied to the giant glitches of the Vela pulsar, for which we can reproduce the approximate waiting time between glitches, the size of the glitch and the short term post-glitch relaxation. The main assumption is that a giant glitch will occur once the system exceeds the maximum lag that the pinning force in the crust can sustain. This naturally gives rise to a waiting time between glitches that depends on the pulsar spindown rate (i.e. it is the time it takes the crust to spin down by the required amount) and to a maximum size for the glitch. Both these quantities depend only weakly on the mass and radius of the star \citep{pinning1, pinning1phd, Stefano} and our model can, in fact, reproduce these features successfully also for the general population of "giant glitchers", i.e. the pulsars for which giant glitches such as those of the Vela have been observed \citep{Espinoza}.

In our model the coupling between the charged component (which we assume to be rigidly rotating) and the superfluid neutrons is given by the vortex-mediated mutual friction.  Our results suggest that the mutual friction will be weak in the crust, possibly due to the fact that not all vortices are free, but rather that the strong pinning force gives rise to a situation in which most vortices are pinned and only a small fraction can 'creep' outwards. Only once the maximum unpinning lag is exceeded can the vortices move out freely; a process which can excite Kelvin oscillations and give rise to a strong drag and recoupling of the two components on a very short timescale, i.e. a glitch. The short term post-glitch relaxation of the Vela, on the other hand, suggests that the magnitude of the drag in the core of the NS is consistent with theoretical expectations for electron scattering of magnetised vortex cores. Our model does not support the notion that, at least on short timescales, a significant number of vortices is pinned in the core (as could, for example, be the case if one has a type II superconductor and vortices cannot cross fluxtubes, effectively decoupling the core and the crust). A detailed analysis of the case in which the core consists of a type II superconductor will be a focus of future work in order to obtain more quantitative results and constraints on NS interior physics.  Some vortices that cross the core  may however be weakly pinned to the crust, and vortex repinning and creep (also in the core) may play a role on the longer timescales associated with the recovery.

Another effect which will have an impact on the post-glitch recovery is the Ekman flow at the crust-core interface. This effect has been shown to be important in fitting the post glitch recovery of the Vela and Crab pulsars by \citet{vE1} and future adaptations of our model should relax the rigid rotation assumption for the charged component and include the effect of Ekman pumping.  Further developments should also include more realistic models for the drag parameters in the star, as the density dependence of the coupling strength clearly has an impact on the amount of angular momentum that can be exchanged on different timescales. Truly quantitative results could then be obtained with the use of realistic equations of state together with consistent estimates of the pinning force, such as those of \citep{pinning1} and \citep{pinning1phd}.

Note that we have assumed that a giant glitch only occurs when the maximum critical lag is reached. If unpinning could be triggered earlier, this could generate smaller glitches. In fact cellular automaton models have shown that the waiting time and size distributions of pulsar glitches can be successfully explained by vortex avalanche dynamics, related to random unpinning events \citep{Melatos1, Melatos2,NewM}. It would thus be of great interest to use our long-term hydrodynamical models, with realistic pinning forces, as a background for such cellular automaton models that model the short-term vortex dynamics. Such a model could then also be extended to model not only large pulsar glitches, but more generally pulsar timing noise, an issue that is of great importance for the current efforts to detects GWs with pulsar timing arrays \citep{Hobbs}.

Finally, the next generation of radio telescopes, such as LOFAR and the SKA, is likely to provide much more precise timing data for radio pulsars and is likely to set much more stringent constraints on the glitch rise time and short term relaxation, thus allowing us to test our models and probe the coupling between the interior superfluid and the crust of the NS with unprecedented precision.

\section*{Acknowledgments}

This work was supported by CompStar, a Research Networking Programme of the European Science Foundation.

BH would like to thank Cristobal Espinoza, Danai Antonopoulou and Fabrizio Grill for stimulating discussions on pulsar glitch observations and pinning force calculation. BH also acknowledges support from the European Union via a Marie-Curie IEF fellowship and from the European Science Foundation (ESF) for the activity entitled "The New Physics of Compact Stars" (COMPSTAR) under exchange grant 2449.
  
TS acknowledges support from EU FP6 Transfer of Knowledge project ``Astrophysics of Neutron Stars'' (ASTRONS, MTKD-CT-2006-042722).

\end{document}